\documentclass[12pt,english]{article}
\usepackage[T1]{fontenc}
\usepackage[latin9]{inputenc}
\usepackage{geometry}
\usepackage{amstext}
\usepackage{graphicx}
\usepackage{setspace}
\usepackage{esint}
\makeatletter

\providecommand{\tabularnewline}{\\}

\@ifundefined{date}{}{\date{}}
\makeatother

\usepackage{babel}
\begin{document}

\title{\textbf{\Large{}A Mechanical Approach to One-Dimensional Interacting
Gas}}

\author{{\normalsize{}Chung-Yang Wang and Yih-Yuh Chen}\\
\emph{\normalsize{}Department of Physics, National Taiwan University,
Taipei 106, Taiwan, Republic of China}}
\maketitle
\begin{abstract}
Traditional derivations of the van der Waals equation typically use
standard recipes involving ensemble averages of statistical mechanics.
In this work, we study a box of weakly interacting gas particles in
one-dimension from a purely mechanical point of view. This has the
merit that it not only reproduces the van der Waals equation but also
tells us some extra interesting physics not immediately clear from
a pure statistical mechanical approach. For example, we find that
the traditional handwaving interpretation of the van der Waals equation
adopting mean field approximation is actually incorrect. In this investigation
of one-dimensional interacting gas, we demonstrate the possibility
taking a mechanical point of view and having deeper understanding
for the physics of leading order effect of particle-particle interaction,
for weakly interacting N-body systems that are usually studied in
the framework of statistical mechanics or kinetic theory.

(This paper is the simplified version of the master thesis of Chung-Yang
Wang at National Taiwan University.)
\end{abstract}
\medskip{}

\medskip{}
\medskip{}
\medskip{}

\medskip{}

\medskip{}
\medskip{}

\medskip{}
\medskip{}

\medskip{}
\medskip{}

\medskip{}
\medskip{}

\medskip{}
\medskip{}

\medskip{}
\medskip{}
\medskip{}

\medskip{}

\medskip{}
\medskip{}

\medskip{}
\medskip{}

\medskip{}
\medskip{}
\medskip{}
\medskip{}

\medskip{}
\medskip{}
E-mail: chyawang@terpmail.umd.edu

\pagebreak{}

\section{Introduction}

Statistical mechanics and kinetic theory are two main frameworks one
adopts when studying a system containing a huge number of degrees
of freedom. Researches in this field concern various systems, including
hard sphere system \cite{dufty1996practical,rohrmann2007structure,tonks1936complete,uranagase2007effects,uranagase2006statistical,urrutia2008two,visco2008collisional,wang2002van},
hard ellipse system \cite{foulaadvand2013two}, hard needle system
\cite{gurin2011towards,kantor2009one}, van der Waals theory \cite{largo2003generalized,barra2015exact,giglio2016integrable,zhong2017modified},
granular particle system \cite{brey2005hydrodynamic,goldhirsch2005nearly,risso2002dynamics},
etc. There are also researches concerning fundamental issues \cite{liu2014investigation,mazenko2010fundamental}.

Statistical mechanics and kinetic theory are powerful and systematic,
but it often amounts to meaning that one has to pay the price of losing
certain detailed dynamical information about the interparticular interactions.
In order to study a weakly interacting N-body system, besides statistical
mechanics, kinetic theory and numerical simulation, do we have other
choice? At first glance, the idea of \emph{particle trajectory} is
useful only for systems containing small degrees of freedom. However,
we can actually bring in the idea of trajectory for a N-body system
if the effect of particle-particle interaction is weak. In this work,
we study a weakly interacting one-dimensional gas to see how a direct
approach investigating the detailed mechanical interaction between
particles might shed some light on its merits as opposed to the traditional
approach.

Before describing our mechanical approach, we first give a brief review
of how one understands a weakly interacting one-dimensional gas in
statistical mechanics. For a one-dimensional classical gas in the
box, the equation of state at thermal equilibrium can be calculated
by standard recipes in statistical mechanics \cite{feynman1998statistical,salinas2013introduction},
and is given by 

\begin{equation}
F=\rho(k_{B}T_{real})-\rho^{2}(k_{B}T_{real})\int_{0}^{\infty}dr\left[e^{-\frac{U(r)}{k_{B}T_{real}}}-1\right]+\mathcal{O}(\rho^{3}),\label{eq:1-1}
\end{equation}
where $F$ is the force exerted on each side of the confining box,
$\rho\equiv N/L$ is the linear number density ($N$ and $L$ are
the particle number and the size of the one-dimensional box, respectively),
$k_{B}$ is Boltzmann's constant, $T_{real}$ is the temperature of
the system (It will be clear why we use the notation $T_{real}$ instead
of $T$.), and $U(r)$ is the potential between two gas particles
with $r$ being their separation. In the above, one has assumed a
low density limit so that a Taylor expansion in $\rho$ is possible.

We will specialize to the case when the potential representing particle-particle
interaction is consisted of a hard core and an interaction tail, as
shown in Fig. 1.

\begin{figure}
\begin{centering}
\includegraphics[scale=0.33]{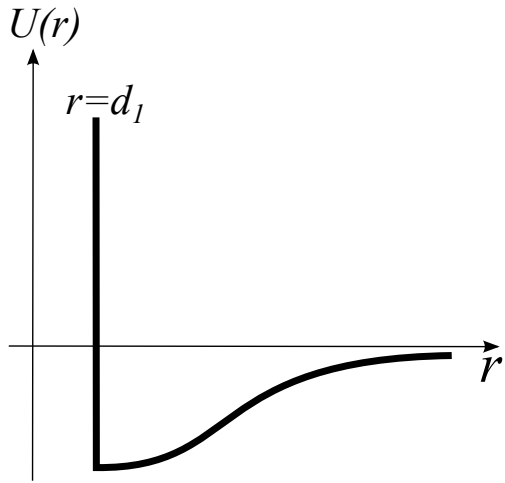}
\par\end{centering}
\caption{Particle-particle interaction consisting of a hard core (at $r=d_{1}$)
and an interaction tail. $U(r)$ is the potential describing the interaction
between two particles.}

\end{figure}

Consider only the low density limit and keep up to the second order
in the number density, and further assume the high temperature limit
and weakly interacting limit. Approximate Eq.\ref{eq:1-1} to first
order of $U(r)/k_{B}T_{real}$ under these limits, and the equation
of state becomes

\begin{equation}
F=\rho(k_{B}T_{real})\left(1+\rho d_{1}-\rho\int_{d_{1}}^{\infty}dr\left[\frac{-U(r)}{k_{B}T_{real}}\right]\right).\label{eq:1-3}
\end{equation}
After rearrangement, the equation of state becomes

\begin{equation}
\left(F+(\frac{N}{L})^{2}\int_{d_{1}}^{\infty}dr\left[-U(r)\right]\right)\left(L-d_{1}N\right)=Nk_{B}T_{real},\label{eq:1-4}
\end{equation}
which is the van der Waals equation in one-dimension.

This is the statistical mechanics approach to the equation of state
of interacting gas under the conditions of low density, high temperature
and weak interaction. The derivation provided by statistical mechanics
is simple, and in one scoop it easily relates the force with the temperature,
the density and the potential describing particle-particle interaction.
On the other hand, one also loses track of exactly what has happened
to the \emph{dynamical} behavior of the constituent particles. For
example, if we ``turn on\textquotedblright{} the interparticular
interactions of an otherwise ideal gas, will that make the gas particle
move faster or slower? And, does the answer depend on the original
velocity of a particle? Statistical mechanics by itself doesn't provide
such information. In this regard, the physics contained in an equation
of state is quite limited.

In this work, we adopt an approach different from statistical mechanics.
We take a mechanical point of view, considering trajectories and detailed
mechanics of the gas particles. Besides rebuilding the equation of
state, we will have some physics that standard statistical mechanics
doesn't tell us. For example, we will try to answer the two previous
questions in our mechanical approach.

The paper is organized as follows. In Section 2, we discuss the foundation
of our mechanical model. In Section 3, we consider a special case
of square well particle-particle interaction and show how the mechanical
model is built. In Section 4, we generalize the previous results to
generic interacting gas, with the help of physics insight we developed
in Section 3. Finally, we summarize the result and give some possible
implications in Section 5.

\section{Mechanical picture for one-dimensional interacting gas}

Consider first a box of \emph{ideal} gas in one-dimension. For visualization,
we attach a velocity arrow to each particle. When two particles collide,
they will exchange their velocity arrows. Thus, we may follow an arrow
all the way as it propagates from left to right, totally ignoring
the fact that it is actually the different particles that are carrying
it at different times. To avoid possible misunderstanding, we want
to emphasize that we are tracking a velocity arrow instead of a particle
itself. The time-averaged force experienced by the wall is then given
by

\begin{equation}
F=\sum_{j}\frac{2mv_{j}^{0}}{T_{j}^{0}},\label{eq:2-3}
\end{equation}
where $j$ is the index of each arrow hitting the wall, $v_{j}^{0}$
being the velocity associated with the $j$-th arrow, and $T_{j}^{0}=2L/v_{j}^{0}$
is the total time it takes for a specific arrow to complete one circuit.
The superscript $0$ stands for free particle (the unperturbed state).

\begin{figure}
\begin{centering}
\includegraphics[scale=0.7]{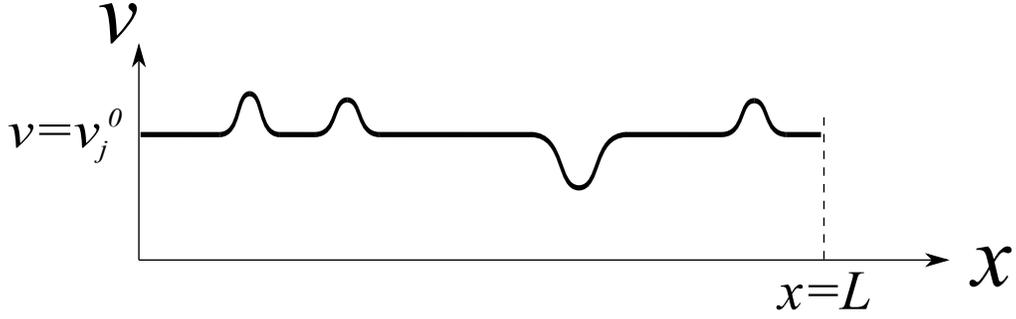}
\par\end{centering}
\caption{Velocity fluctuation of a particle (actually, a velocity arrow) for
a typical motion.}

\end{figure}

Summation over $j$ is given by

\begin{equation}
\sum_{j}(...)=\frac{2N}{\sqrt{a\pi}}\int_{0}^{\infty}dv_{j}^{0}e^{-\frac{\left(v_{j}^{0}\right)^{2}}{a}}(...),\label{eq:2-5}
\end{equation}
where $a\equiv2k_{B}T/m$. Define dimensionless velocity $u\equiv v/\sqrt{a}$,
which will be used in later calculations. Plugging Eq.\ref{eq:2-5}
into Eq.\ref{eq:2-3}, we have
\begin{equation}
F=\rho k_{B}T,\label{eq:2-6}
\end{equation}
which is the one-dimensional version for $PV=Nk_{B}T$.

When particle-particle interaction is turned on, a particle is no
longer a free particle and its velocity fluctuates, as shown in Fig.
2. Symbolically, we have

\begin{equation}
F=\sum_{j}\frac{2mv_{j}}{T_{j}}=\sum_{j}\frac{2m(v_{j}^{0}+\Delta v_{j}^{0})}{T_{j}^{0}+\Delta T_{j}^{0}},\label{eq:2-7}
\end{equation}
where $v_{j}^{0}$ is the free particle velocity, $T_{j}^{0}$ is
the flying time period without interaction, $v_{j}=v_{j}^{0}+\Delta v_{j}^{0}$
and $T_{j}=T_{j}^{0}+\Delta T_{j}^{0}$ are the collision velocity
and the flying time period in the presence of interaction. The question
we want to ask is, when there is particle-particle interaction, how
do flying time period (physics in the bulk) and momentum transferred
(physics on the boundary) change, and hence lead to the change of
equation of state? Note that standard statistical mechanics cannot
tell us ``what happens in the bulk'' and ``what happens on the
boundary.'' It just gives us the equation of state, the net result
after combining ``physics in the bulk'' and ``physics on the boundary.''

In the mechanical approach, we consider the limits of low density
(dilute gas), short-range and weak particle-particle interaction (weakly
interacting) and high temperature. Compared with ideal gas, the correction
due to particle-particle interaction in van der Waals equation of
state is kept to order one, which can be seen in Eq.\ref{eq:1-3}.
Therefore, in order to compare with the van der Waals equation, we
consider the physics perturbed around ideal gas (free particle) regime
and need only retain the perturbation up to the first order correction.

Keeping the physics to the first order correction, Eq.\ref{eq:2-7}
becomes
\begin{equation}
F=\sum_{j}\frac{2mv_{j}^{0}}{T_{j}^{0}}-\sum_{j}\frac{2mv_{j}^{0}}{T_{j}^{0}}\frac{\Delta T_{j}^{0}}{T_{j}^{0}}+\sum_{j}\frac{2m\Delta v_{j}^{0}}{T_{j}^{0}}.\label{eq:2-8}
\end{equation}
In order to do the summation ($\sum_{j}$), we will make the explicit
assumption that Maxwell's velocity distribution is valid for $v_{j}^{0}$.
That is, in our mechanical model, the unperturbed part (free particle
part) described by $v_{j}^{0}$ satisfies Maxwell's velocity distribution
of temperature $T$. But is this $T$ equal to \emph{the} \emph{real
temperature} \emph{of the box of interacting gas}? We don't know at
this stage. The real temperature of the interacting gas (free particle
part of temperature $T$ plus the effect of particle-particle interaction),
denoted by $T_{real}$, will be studied in Section 3.3. Note that
if we have $T\neq T_{real}$, then we have $\sum_{j}\left(2mv_{j}^{0}/T_{j}^{0}\right)=\rho k_{B}T\neq\rho k_{B}T_{real}$,
which means that it is too naive to identify $\sum_{j}\left(2mv_{j}^{0}/T_{j}^{0}\right)$
(in our mechanical approach) with the ideal gas part in Eq.\ref{eq:1-1}
to Eq.\ref{eq:1-4} (in standard statistical mechanics). Here we want
to stress that temperature is a thermodynamic concept instead of a
mechanical concept, and hence calculations are well-defined in our
\emph{mechanical approach}, without dealing with the \emph{thermodynamic
issue} of ``What is the real temperature of the system?''

To avoid possible confusion, we should emphasize that we are talking
about the real temperature of the system when taking standard statistical
mechanics. This is why we use $T_{real}$ in Eq.\ref{eq:1-1} to Eq.\ref{eq:1-4}.

What we are going to do with our mechanical model is trying to study
the three terms in Eq.\ref{eq:2-8} with the notion of particle trajectories.
The first term corresponds to ideal gas, namely the unperturbed part.
The second term arises from the modification of flying time period.
The third term comes from the modification of collision velocity.
The strategy is tracking a particle (actually, a velocity arrow) labeled
by $j$ (which means that its free particle velocity is $v_{j}^{0}$),
called the main particle, and finding out the influence of the other
particles, called the background particles, on the main particle.
Apparently, the main particle is not special but just a notion for
bookkeeping.

At the end of this section, we want to point out that interacting
gas is different from ideal gas due to particle-particle interaction.
Therefore, it is not surprising that ``how many particle-particle
interactions are there'' and ``what is the influence of a particle-particle
interaction'' are at the heart of our mechanical approach when studying
interacting gas. We will see how to understand the behavior of a box
of interacting gas by these two central ideas.

\section{One-dimensional interacting gas with square well potential}

In this section, we consider one-dimensional interacting gas with
square well potential, as shown in Fig. 3. By Eq.\ref{eq:1-3}, equation
of state for square well potential is

\begin{equation}
F=\rho\left(k_{B}T_{real}\right)+\rho^{2}\left(k_{B}T_{real}\right)d_{1}-\rho^{2}\left(d_{2}-d_{1}\right)\epsilon.\label{eq:A-1}
\end{equation}
Eq.\ref{eq:A-1} is what we are going to compare with when obtaining
an equation of state by our mechanical approach.

\begin{figure}
\begin{centering}
\includegraphics[scale=0.4]{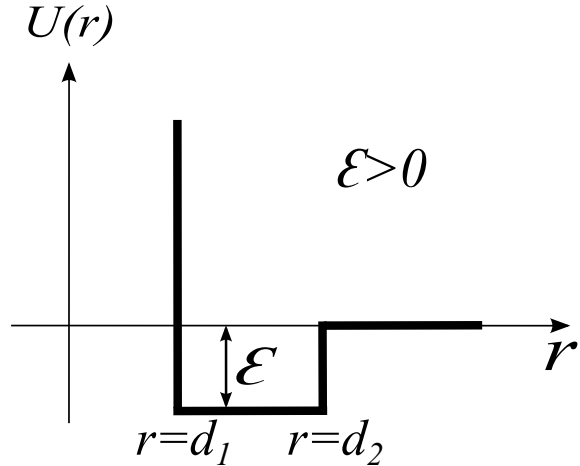}
\par\end{centering}
\caption{Square well potential.}

\end{figure}

\subsection{Mechanics of interaction between two particles}

Consider generic particle-particle interaction with a finite interaction
range $r_{0}$. Suppose the interaction takes time interval of $\Delta t_{interaction}$.
In the lab frame (of the confining box), the displacements of two
interacting particles during $\Delta t_{interaction}$ are given by

\begin{equation}
\mathrm{displacement\ of\ left\ particle}=r_{0}+v_{CM}\Delta t_{interaction}\label{eq:A-2}
\end{equation}
and

\begin{equation}
\mathrm{displacement\ of\ right\ particle}=-r_{0}+v_{CM}\Delta t_{interaction},\label{eq:A-3}
\end{equation}
where $v_{CM}$ is the center-of-mass velocity. For a potential shown
in Fig. 1, $\Delta t_{interaction}\left(v_{1},v_{2}\right)$ is given
by

\begin{equation}
\Delta t_{interaction}\left(v_{1},v_{2}\right)=\int_{d_{1}}^{r_{0}}\frac{dr}{\sqrt{\left(\frac{v_{1}-v_{2}}{2}\right)^{2}-\frac{U(r)}{m}}}.\label{eq:A-5}
\end{equation}

For particle-particle repulsion, $\Delta t_{interaction}$ is more
complicated in the sense that the lower bound of $d_{1}$ of the integral
in Eq.\ref{eq:A-5} should be replaced by a function of $v_{1}$ and
$v_{2}$. Particle-particle repulsion will be discussed later in Section
4.2.

\subsection{Flying time period (physics in the bulk)}

Without loss of generality, we may track a main ``particle\textquotedblright{}
(in fact, a velocity arrow) which flies from the left wall to the
right wall. How is the flying time period modified by particle-particle
interaction? We first study number of collisions for the main particle
in the trip flying from the left wall to the right wall (Section 3.2.1).
This number turns out \emph{not} to have anything to do with the form
of the particle-particle interaction, as we will show in the following.
The details of the particle-particle interaction comes in only when
we are dealing with the effect of each collision (Section 3.2.2).

Whenever the main particle is moving outside the interaction range
of the rest of the particles (the ``background\textquotedblright ),
it regains its free particle velocity $v_{j}^{0}$. Thus, the flying
time period is given by
\begin{equation}
\frac{1}{2}T_{j}=\frac{L-\mathrm{total\ displacement\ of\ }v_{j}^{0}\ \mathrm{in\ interaction}}{v_{j}^{0}}+\sum_{k}\Delta t_{interaction}(v_{j}^{0},v_{k}),\label{eq:B-1}
\end{equation}
where the background particles are labeled by $k$, and $v_{k}$ is
the velocity of the background particle when colliding with the main
particle. Particle-particle interactions can be classified as forward
collision and backward collision. By forward collision, we mean the
background particle collides with the main particle from the right.
By backward collision, we mean the background particle collides with
the main particle from the left. The numbers of forward collisions
and backward collisions of the main particle in its whole trip are
denoted as $\#_{R}$ and $\#_{L}$, respectively. Putting Eq.\ref{eq:A-2}
and Eq.\ref{eq:A-3} into Eq.\ref{eq:B-1}, we have

\begin{eqnarray}
\frac{1}{2}\Delta T_{j}^{0} & = & \#_{R}\text{·}\left(-\frac{r_{0}}{v_{j}^{0}}+\frac{v_{j}^{0}-v_{k}}{2v_{j}^{0}}\Delta t_{interaction}(v_{j}^{0},v_{k})\right)\nonumber \\
 &  & +\#_{L}\text{·}\left(\frac{r_{0}}{v_{j}^{0}}+\frac{v_{j}^{0}-v_{k}}{2v_{j}^{0}}\Delta t_{interaction}(v_{j}^{0},v_{k})\right).\label{eq:B-4}
\end{eqnarray}
(Recall that $\Delta T_{j}^{0}=T_{j}-T_{j}^{0}$.) So we need to find
out $\#_{R}$, $\#_{L}$, $v_{k}$ and $\Delta t_{interaction}(v_{j}^{0},v_{k})$.
Now we are going to find out $\#_{R}$ and $\#_{L}$.

\subsubsection{Counting the number of collisions}

For a box of ideal gas, how many collisions does the main particle
experience in the whole trip? For a background particle with specified
initial velocity ($v_{k}^{0}$) and initial position ($x_{k}^{0}$),
the numbers of forward collisions and backward collisions provided
by the background particle are completely determined, as shown in
Fig. 4.

\begin{figure}
\begin{centering}
\includegraphics[scale=0.45]{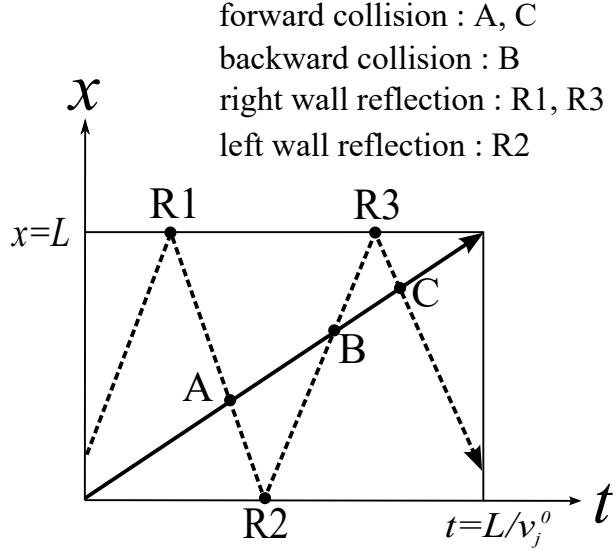}
\par\end{centering}
\caption{The $x-t$ diagram for a typical motion of the main particle (solid
line) and a background particle (dashed line).}

\end{figure}

Ideal gas is easy to study. But what about interacting gas particles?
Because we are only computing corrections correct to the first order
of particle-particle interaction, it turns out that the counting of
the number of particle collisions can be essentially done by assuming
that the particles behave just like ideal gas particles! This is the
key point of our mechanical model. Mathematically speaking, consider
a function $A(\delta)=B(\delta)C(\delta)$, where $\delta$ is small.
Expand $A(\delta)$ with $\delta$

\begin{equation}
A(\delta)=\left(B_{0}+\delta B_{1}+\mathcal{O}(\delta^{2})\right)\left(C_{0}+\delta C_{1}+\mathcal{O}(\delta^{2})\right).\label{eq:100}
\end{equation}
If $B_{0}=0$ and $A(\delta)$ is kept to $\mathcal{O}\left(\delta\right)$,
we have

\begin{equation}
A(\delta)=\left(\delta B_{1}\right)C_{0}.\label{eq:102}
\end{equation}

Total effect of particle-particle interaction, effect of each collision,
number of collisions and strength of particle-particle interaction
play the role of $A(\delta)$, $B(\delta)$, $C(\delta)$ and $\delta$
in Eq.\ref{eq:100} and Eq.\ref{eq:102}, respectively. When the effect
of particle-particle interaction is kept to order one, in order to
count the number of collisions, the free particle model (ideal gas)
is sufficient.

With some elementary counting of the intersection points of straight
lines (For example, there are two forward collisions and one backward
collision in Fig. 4.), $\frac{1}{L}\int_{0}^{L}dx_{k}^{0}\#_{L}\left(v_{j}^{0},v_{k}^{0},x_{k}^{0}\right)$
is calculated and the result is shown in Fig. 5. For $\#_{R}$, we
have

\begin{equation}
\frac{1}{L}\int_{0}^{L}dx_{k}^{0}\#_{R}\left(v_{j}^{0},v_{k}^{0},x_{k}^{0}\right)=1+\frac{1}{L}\int_{0}^{L}dx_{k}^{0}\#_{L}\left(v_{j}^{0},v_{k}^{0},x_{k}^{0}\right).\label{eq:B-8}
\end{equation}

\begin{figure}
\begin{centering}
\includegraphics[scale=0.65]{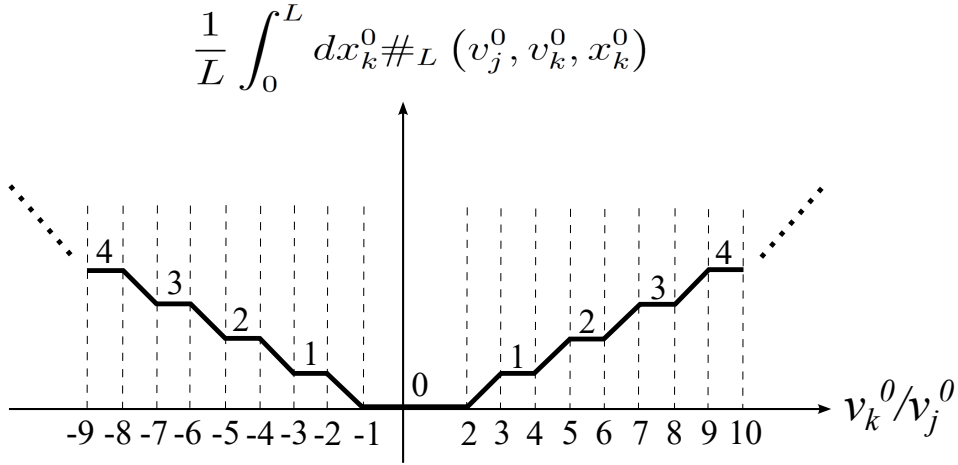}
\par\end{centering}
\caption{$\frac{1}{L}\int_{0}^{L}dx_{k}^{0}\#_{L}\left(v_{j}^{0},v_{k}^{0},x_{k}^{0}\right)$
as a function of $v_{k}^{0}/v_{j}^{0}$. The line segments connecting
two adjacent flat parts are straight lines with slopes being $\pm1$.}

\end{figure}

Let's see what do Fig. 5 and Eq.\ref{eq:B-8} mean. To have a backward
collision with the main particle, a background particle needs to fly
fast enough to catch up with the main particle, which can be seen
from $\frac{1}{L}\int_{0}^{L}dx_{k}^{0}\#_{L}\left(v_{j}^{0},v_{k}^{0},x_{k}^{0}\right)=0$
for $-1<\left(v_{k}^{0}/v_{j}^{0}\right)<2$ in Fig. 5. Comparing
the initial situation at $t=0$ and the final situation at $t=T_{j}^{0}/2$,
all background particles change from sitting at the right side of
the main particle to its left side. Therefore, the number of forward
collisions is larger than the number of backward collisions by one,
as shown in Eq.\ref{eq:B-8}.

Another viewpoint for the number of collisions is considering the
condition of neighbor particles surrounding the main particle to collide
with the main particle. The admissible velocities of neighbor particles
causing forward collisions and backward collisions are $\left(-\infty,v_{j}^{0}\right)$
and $\left(v_{j}^{0},\infty\right)$, respectively, and hence forward
collisions are more than backward collisions.

For a given $\left(x_{k}^{0},v_{k}^{0}\right)$, the corresponding
$v_{k}$ can be found, with the help of $x-t$ diagram. Substituting
the position distribution (homogeneous), the velocity distribution
(Maxwellian), $\#_{R}$, $\#_{L}$ and $v_{k}$ into Eq.\ref{eq:B-4},
we have

\begin{equation}
\frac{1}{2}\Delta T_{j}^{0}=-r_{0}\frac{N}{v_{j}^{0}}+\frac{N}{2}\frac{1}{\sqrt{a\pi}\left(v_{j}^{0}\right)^{2}}\int_{0}^{\infty}dv_{k}^{0}\left[e^{-\frac{\left(v_{k}^{0}-v_{j}^{0}\right)^{2}}{a}}-e^{-\frac{\left(v_{k}^{0}+v_{j}^{0}\right)^{2}}{a}}\right]\left(v_{k}^{0}\right)^{2}\Delta t_{interaction}\left(0,v_{k}^{0}\right).\label{eq:B-17}
\end{equation}

\subsubsection{Correction of flying time period}

Obtain $\Delta t_{interaction}\left(0,v_{k}^{0}\right)$ by Eq.\ref{eq:A-5}
for square well potential and then put it into Eq.\ref{eq:B-17}.
Expand with $\frac{\epsilon}{k_{B}T}$ to $\mathcal{O}\left(\frac{\epsilon}{k_{B}T}\right)$.
We have

\begin{equation}
\frac{1}{2}\Delta T_{j}^{0}=-\frac{Nd_{1}}{v_{j}^{0}}-\frac{N\left(d_{2}-d_{1}\right)\sqrt{a\pi}}{\left(v_{j}^{0}\right)^{2}}e^{-\frac{\left(v_{j}^{0}\right)^{2}}{a}}\mathrm{erfi}(\frac{v_{j}^{0}}{\sqrt{a}})\frac{\epsilon}{k_{B}T}.\label{eq:B-23}
\end{equation}
This is the flying time period correction for square well potential.

Let's try to figure out the physics behind Eq.\ref{eq:B-23}. The
effect of hard core is decreasing the space that a particle can move,
and the corresponding flying time correction is $\frac{1}{2}\Delta T_{j}^{0}=-Nd_{1}/v_{j}^{0}$.
The second term is the effect of the attraction potential. It is always
negative, which means that all particles are sped up. But why? The
physics can be analyzed by average velocity $v_{j}\mid_{average}=\frac{2\left(L-Nd_{1}\right)}{T_{j}^{0}+\Delta T_{j}^{0}}$.
By Eq.\ref{eq:B-23}, we get

\begin{equation}
\frac{u_{j}\mid_{average}-u_{j}^{0}}{\rho\left(d_{2}-d_{1}\right)\frac{\epsilon}{k_{B}T}}=\sqrt{\pi}e^{-\left(u_{j}^{0}\right)^{2}}\mathrm{erfi}(u_{j}^{0}).\label{eq:B-26}
\end{equation}
(Recall that $u\equiv v/\sqrt{a}$.) The result is shown in Fig. 6.

\begin{figure}
\begin{centering}
\includegraphics[scale=0.4]{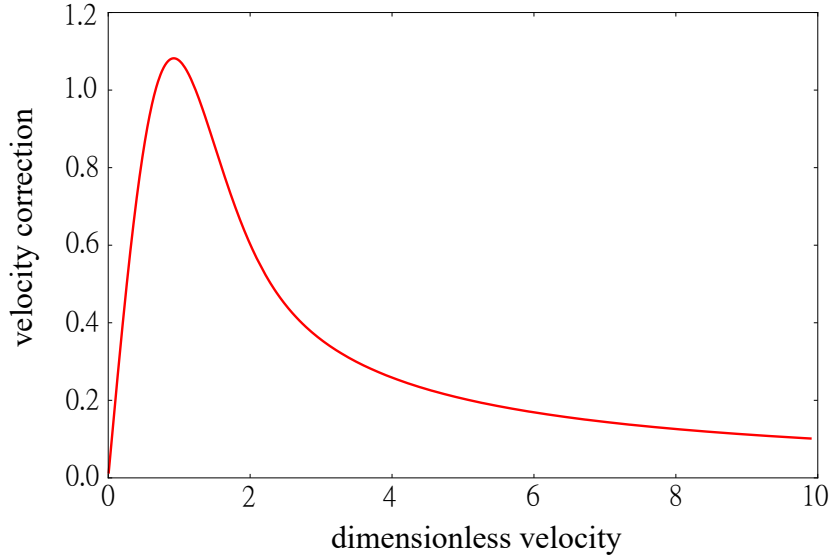}
\par\end{centering}
\caption{Correction of average velocity due to particle-particle attraction
for square well potential. The x-axis is $u_{j}^{0}$ and the y-axis
is $\frac{u_{j}\mid_{average}-u_{j}^{0}}{\rho\left(d_{2}-d_{1}\right)\frac{\epsilon}{k_{B}T}}$
given by Eq.\ref{eq:B-26}.}
\end{figure}

In Fig. 6, we have three features: $u_{j}\mid_{average}-u_{j}^{0}>0$,
an increasing trend and a decreasing trend. In the following, we analyze
the three features by number of collisions and effect of each collision.

For particle-particle attraction, when the main particle has a forward
collision, its velocity is increased; when the main particle has a
backward collision, its velocity is decreased. Since forward collisions
are more than backward collisions, a particle is sped up when particle-particle
attraction is turned on.

By Eq.\ref{eq:B-8} and Fig. 5, the number of collisions provided
by a background particle (with position distribution having been averaged
by $\frac{1}{L}\int_{0}^{L}dx_{k}^{0}$) can be written as
\begin{equation}
1\ \mathrm{forward\ collision}+n\ \left(\mathrm{forward\ collision},\mathrm{backward\ collision}\right)\ \mathrm{pair},\label{eq:B-30}
\end{equation}
where $n=\frac{1}{L}\int_{0}^{L}dx_{k}^{0}\#_{L}\left(v_{j}^{0},v_{k}^{0},x_{k}^{0}\right)$.
For a background particle, the effect of a $(\mathrm{forward\ collision},$
$\mathrm{backward\ collision})$ pair is decreasing the velocity of
the main particle, since the effect of particle-particle interaction
is significant when the relative velocity is small. The feature that
nonvanishing $n$ decreases as $v_{j}^{0}$ increases leads to the
increasing trend in Fig. 6.

Velocity domain of background particles having significant contributions
is a small neighborhood around $v_{j}^{0}$. When $v_{j}^{0}$ increases,
the velocity of this small neighborhood also increases, and hence
the population in this domain decreases, due to the exponential decay
of Maxwell's velocity distribution. This is the reason for the decreasing
trend.

In the traditional picture of an interacting gas, particle-particle
interaction disappears for the bulk part under mean field approximation
\cite{van2004continuity}. With our mechanical approach, now we know
that such handwaving picture is actually incorrect. Background particles
are uniform, but the effect is not zero, since the main particle flies
to a certain direction and thus breaks forward-backward symmetry.
(The admissible velocities of neighbor particles causing forward collisions
and backward collisions are $\left(-\infty,v_{j}^{0}\right)$ and
$\left(v_{j}^{0},\infty\right)$, respectively.) The effect of forward
collisions and the effect of backward collisions don't cancel out.
A particle is sped up when particle-particle attraction is turned
on.

\subsection{Temperature modification}

Now let's deal with the thermodynamic issue of ``What is the real
temperature of the system?'' In Section 3.2, we see that particles
fly faster when particle-particle attraction is turned on. Consequently,
the temperature $T_{real}$ of the box of interacting gas is greater
than it was ($T$) before the interaction is turned on. What this
means is that, one really should start with an ideal gas with a suitably
chosen lower temperature to perform the perturbation calculation,
so that in the end the combined effect of the added interaction will
raise the temperature of the system to the desired temperature. So
what is the relation between $T_{real}$ and $T$?

The idea is, for average kinetic energy $\left\langle K.E.\right\rangle $,
we should get $N\left\langle K.E.\right\rangle =\frac{N}{2}k_{B}T_{real}$.
That is, the real temperature should be introduced by

\begin{equation}
\sum_{j}\frac{1}{2}m\left(v_{j}\mid_{average}\right)^{2}=\frac{N}{2}k_{B}T_{real}.\label{eq:C-3}
\end{equation}
Putting $v_{j}\mid_{average}=\frac{2\left(L-Nd_{1}\right)}{T_{j}^{0}+\Delta T_{j}^{0}}$
and $\sum_{j}\frac{1}{2}m\left(v_{j}^{0}\right)^{2}=\frac{N}{2}k_{B}T$
into Eq.\ref{eq:C-3}, we get

\begin{equation}
\frac{N}{2}k_{B}T_{real}=\frac{N}{2}k_{B}T\left(1-2\rho d_{1}\right)-\sum_{j}m\left(v_{j}^{0}\right)^{2}\frac{\Delta T_{j}^{0}}{T_{j}^{0}}.\label{eq:C-6}
\end{equation}
Eq.\ref{eq:C-6} is the relation between $T_{real}$ and $T$. Putting
Eq.\ref{eq:C-6} into Eq.\ref{eq:2-8}, the equation of state is modified
to

\begin{equation}
F=\rho k_{B}T_{real}+2\rho^{2}\left(k_{B}T_{real}\right)d_{1}+\sum_{j}\frac{2mv_{j}^{0}}{T_{j}^{0}}\frac{\Delta T_{j}^{0}}{T_{j}^{0}}+\sum_{j}\frac{2m\Delta v_{j}^{0}}{T_{j}^{0}}.\label{eq:C-9}
\end{equation}

Actually, the idea of temperature modification can also be seen by
considering the total energy of the box of gas. In our current formulation,
turning on the interaction does not change the total energy of the
system. Thus, on the one hand, this total energy should be equal to
the total kinetic energy of the unperturbed ideal gas, which implies

\begin{equation}
\mathrm{total}\ \mathrm{energy}=\sum_{j}\frac{1}{2}m\left(v_{j}^{0}\right)^{2}=\frac{N}{2}k_{B}T.\label{eq:C-11}
\end{equation}
On the other hand, this total energy becomes the summation of the
average kinetic energy and average potential energy of each particle,
which reads

\begin{equation}
\mathrm{total}\ \mathrm{energy}=\frac{N}{2}k_{B}T_{real}+N\left\langle U\right\rangle .\label{eq:C-12}
\end{equation}
Comparing the above two expressions for the same total energy, we
have

\begin{equation}
k_{B}T_{real}=k_{B}T-2\left\langle U\right\rangle .\label{eq:C-13}
\end{equation}
Therefore, $T_{real}>T$ when particle-particle attraction is turned
on, just as claimed before.

Note that we can see that traditional interpretation of the van der
Waals equation adopting mean field approximation is incorrect simply
by $\sum_{j}\frac{1}{2}m\left(v_{j}^{0}\right)^{2}=\sum_{j}\frac{1}{2}m\left(v_{j}\mid_{average}\right)^{2}+N\left\langle U\right\rangle $,
without doing detailed calculations. When particle-particle attraction
is turned on, potential energy is negative and hence particles move
faster.

\subsection{Momentum transferred (physics on the boundary)}

Collision velocity of the main particle is determined by its state
hitting the right wall. How to find out the state of the main particle
at $t=T_{j}^{0}/2$? The idea is, the main particle experiences lots
of collisions during its entire trip, and the last one determines
its behavior at $t=T_{j}^{0}/2$. If the position of the last collision
of the main particle is close to the right wall such that particle-wall
collision is simultaneously happening when two particles are interacting,
nonvanishing $\Delta v_{j}^{0}$ appears. That is, nonvanishing $\Delta v_{j}^{0}$
appears once the wall comes into play during the interaction and thus
interrupts the otherwise ``normal'' interaction.

In order to deal with the effect of $\Delta v_{j}^{0}$, we collect
the cases having nonvanishing $\Delta v_{j}^{0}$. The effect of $\Delta v_{j}^{0}$
can be decomposed into two parts. The first part is about the probability
of a given situation of the last collision (Section 3.4.1). The second
part is about the effect associated with a given situation of the
last collision (Section 3.4.2). The total effect is the product of
the two parts (Section 3.4.3).

\subsubsection{Probability of the last collision}

The setting of the start of the last collision is that the main particle
is at $L-d_{3}$ and the background particle that causes the last
collision hits the main particle with $v_{k}^{0}$. The probability
of such scenario is studied in Appendix A, and is given by
\begin{equation}
N\left(\left\vert 1-\frac{v_{k}^{0}}{v_{j}^{0}}\right\vert \frac{\Delta d_{3}}{L}\right)\left(f\left(v_{k}^{0}\right)\Delta v_{k}^{0}\right)\mathrm{exp}\left\{ -\frac{1}{2}\left[1+\mathrm{erf}\left(\frac{v_{j}^{0}}{\sqrt{a}}\right)+\frac{\sqrt{a}}{\sqrt{\pi}v_{j}^{0}}e^{-\frac{\left(v_{j}^{0}\right)^{2}}{a}}\right]\rho d_{3}\right\} .\label{eq:D-24}
\end{equation}

\subsubsection{Situation around the wall}

Let's consider non-trivial situations (nonvanishing $\Delta v_{j}^{0}$)
of the last collision of the main particle around the wall. For square
well potential, the relation between $v_{in}$ and $v_{out}$ ($v_{in}$
and $v_{out}$ are the speeds inside and outside the potential well,
respectively, defined in the center-of-mass frame of the two interacting
particles.) is

\begin{equation}
v_{in}=\sqrt{\left(v_{out}\right)^{2}+\frac{\epsilon}{m}}.\label{eq:D-25}
\end{equation}
Particle-particle interaction is easy to analyze in the center-of-mass
frame. To investigate the role played by the wall, lab frame is a
better choice. Therefore, we will switch between center-of-mass frame
and lab frame. Note that collision velocity is defined in the lab
frame. In the following analysis, we separate the situations of the
last collision into forward collision part and backward collision
part.

\paragraph{Forward collision part\protect \\
}

\begin{figure}
\begin{centering}
\includegraphics[scale=0.55]{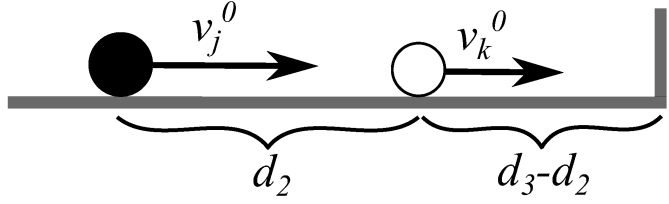}
\par\end{centering}
\caption{The initial state of the last collision being a forward collision.}

\end{figure}
The situation for the beginning of forward collision is shown in Fig.
7. $v_{k}^{0}$ is not arbitrary. First, the occurrence of forward
collision implies $v_{k}^{0}<v_{j}^{0}$. Second, if $v_{CM}=\left(v_{j}^{0}+v_{k}^{0}\right)/2$
is not positive, the two particles don't hit the wall until their
interaction is over. Combining the two considerations, the condition
for $v_{k}^{0}$ is

\begin{equation}
-v_{j}^{0}<v_{k}^{0}<v_{j}^{0}.\label{eq:D-26}
\end{equation}
For the main particle to hit the wall when it is still in the potential
well, $d_{3}$ should be short enough. The range of $d_{3}$ and the
collision velocity is presented in Appendix B.

We put the range of $d_{3}$ and the collision velocity in appendix
because it is the range of $v_{k}^{0}$ that plays the truly significant
role for the feature of physics on the boundary. The role of the range
of $v_{k}^{0}$ will be clear in Section 3.4.3 and Section 4.1.

\paragraph{Backward collision part\protect \\
}

\begin{figure}
\begin{centering}
\includegraphics[scale=0.55]{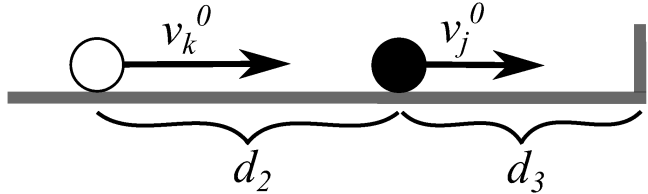}
\par\end{centering}
\caption{The initial state of the last collision being a backward collision.}

\end{figure}
The situation for the beginning of backward collision is shown in
Fig. 8. The occurrence of backward collision implies

\begin{equation}
v_{k}^{0}>v_{j}^{0}.\label{eq:D-31}
\end{equation}
The range of $d_{3}$ and the collision velocity is presented in Appendix
C.

\subsubsection{Correction of collision velocity}

Having considered all the factors, we are now in a position to actually
calculate the combined result for $\Delta v_{j}^{0}$. The total effect
is the product of ``probability of a given situation of the last
collision'' (Section 3.4.1, Appendix A) and ``effect associated
with a given situation of the last collision'' (Section 3.4.2, Appendix
B and Appendix C). The derivation is in Appendix D, E, and F. Finally,
we get

\begin{table}
\begin{centering}
\begin{tabular}{|c|c|c|}
\hline 
 & in the bulk & on the boundary\tabularnewline
\hline 
forward collision & $-\infty\sim v_{j}^{0}$ & $-v_{j}^{0}\sim v_{j}^{0}$\tabularnewline
\hline 
backward collision & $v_{j}^{0}\sim\infty$ & $v_{j}^{0}\sim\infty$\tabularnewline
\hline 
\end{tabular}
\par\end{centering}
\caption{Velocity domains of neighbor particles (surrounding the main particle)
that cause forward and backward collisions.}
\end{table}

\begin{figure}
\begin{centering}
\includegraphics[scale=0.37]{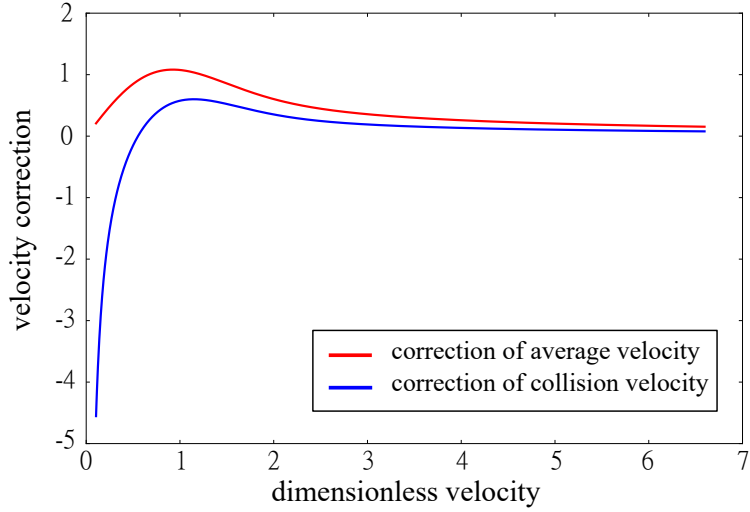}
\par\end{centering}
\caption{Correction of average velocity and correction of collision velocity
for square well potential. The x-axis is $u_{j}^{0}$. The upper curve
is the correction of average velocity given by $\frac{u_{j}\mid_{average}-u_{j}^{0}}{\rho\left(d_{2}-d_{1}\right)\frac{\epsilon}{k_{B}T}}$
in Eq.\ref{eq:B-26}. The lower curve is the correction of collision
velocity given by $\frac{\left(u_{j}^{0}+\Delta u_{j}^{0}\right)-u_{j}^{0}}{\rho\left(d_{2}-d_{1}\right)\frac{\epsilon}{k_{B}T}}$
in Eq.\ref{eq:D-101}.}
\end{figure}

\begin{equation}
\frac{\Delta u_{j}^{0}}{\rho\left(d_{2}-d_{1}\right)\frac{\epsilon}{k_{B}T}}=\frac{1}{2\sqrt{\pi}u_{j}^{0}}\left(\int_{-u_{j}^{0}}^{\infty}du_{k}^{0}e^{-\left(u_{k}^{0}\right)^{2}}\frac{u_{j}^{0}+u_{k}^{0}}{u_{j}^{0}-u_{k}^{0}}+\int_{u_{j}^{0}}^{\infty}du_{k}^{0}e^{-\left(u_{k}^{0}\right)^{2}}\frac{u_{j}^{0}-u_{k}^{0}}{u_{j}^{0}+u_{k}^{0}}\right).\label{eq:D-101}
\end{equation}
The result is shown in Fig. 9.

The profile of $\Delta u_{j}^{0}$ in Fig. 9 can be understood as
we decompose the profile into two main trends, an increasing trend
and a decreasing trend. From Table 1, for physics on the boundary,
as $u_{j}^{0}$ increases, number of relevant forward collisions increases
and number of relevant backward collisions decreases. So $\Delta u_{j}^{0}$
increases as $u_{j}^{0}$ increases. As for the decreasing trend,
the physics is the same as the decreasing trend for $\left(u_{j}\mid_{average}-u_{j}^{0}\right)$.
Analysis here is like what we do for the profile of $\left(u_{j}\mid_{average}-u_{j}^{0}\right)$
in Fig. 6. The underlying physics are both collision number and exponential
decay suppression. This is why $\Delta u_{j}^{0}$ and $\left(u_{j}\mid_{average}-u_{j}^{0}\right)$
have similar profiles in Fig. 9.

We can see that average velocity is always larger than collision velocity.
We will come back to the physical meaning of this feature later when
studying generic particle-particle interaction.

\subsubsection{Correction of momentum transferred in equation of state}

It's time to complete the last mile to a full description of the correction
of momentum transferred in equation of state. Plugging Eq.\ref{eq:D-101}
into $\sum_{j}\left(2m\Delta v_{j}^{0}/T_{j}^{0}\right)$, we get

\begin{equation}
\sum_{j}\frac{2m\Delta v_{j}^{0}}{T_{j}^{0}}=0.\label{eq:D-108}
\end{equation}
In order to obtain $\sum_{j}\left(2m\Delta v_{j}^{0}/T_{j}^{0}\right)$,
we went through all the calculations from Section 3.4.1 to Section
3.4.3 and all the appendices, and yet the final result is just zero!
What's going on? For obtaining such simple result after a long calculation,
we had better give a story. So here is the story.

Note that what we have is $\sum_{j}\left(2m\Delta v_{j}^{0}/T_{j}^{0}\right)=0$
but not $\left(2m\Delta v_{j}^{0}/T_{j}^{0}\right)=0$. A summation
giving zero often relates to conservation law. This is the first hint
for the physics behind Eq.\ref{eq:D-108}.

In our construction, $v_{j}^{0}+\Delta v_{j}^{0}$ is the collision
velocity that a particle hits the wall. A different viewpoint is that
the wall is a momentum provider, since it provides some momentum to
the box of gas every time a particle hits it. Let's check the total
positive momentum of the system. For a box of $N$ particles, in order
to have an idea of total positive momentum that is separate from total
negative momentum, we take a snapshot for the box of gas and pull
apart all particle pairs, and then define total positive momentum
and total negative momentum by these $N$ decoupled particles.

When the box of gas is at thermal equilibrium, it is a steady state,
and hence the total positive momentum is time-independent. After the
system evolves for $\Delta t$, what is the total positive momentum
now? For a particle with $v_{j}^{0}>0$ having initial position in
$\left(L-v_{j}^{0}\Delta t,L\right)$, its velocity will be negative
at $t=\Delta t$. When counting the total positive momentum at $t=\Delta t$,
such particle has no contribution. Such particles have the population
of $\left(v_{j}^{0}\Delta t\right)/L=\Delta t/\frac{1}{2}T_{j}^{0}$.
When counting the total positive momentum at $t=\Delta t$, momentum
loss due to the effect that a particle can change its direction is
given by

\begin{equation}
\mathrm{momentum\ loss}=\sum_{j}\frac{\Delta t}{\frac{1}{2}T_{j}^{0}}\left(-mv_{j}^{0}\right).\label{eq:D-111}
\end{equation}
On the other hand, we also have momentum gain due to the reflection
provided by the left wall. For each collision, the left wall provides
a momentum of $m\left(v_{j}^{0}+\Delta v_{j}^{0}\right)$ to the total
positive momentum of the system. Momentum gain provided by the left
wall during $\Delta t$ is

\begin{equation}
\mathrm{momentum\ gain}=\sum_{j}\frac{\Delta t}{\frac{1}{2}T_{j}^{0}}\left(m\left(v_{j}^{0}+\Delta v_{j}^{0}\right)\right).\label{eq:D-112}
\end{equation}
Since the total positive momentum of the system is time-independent,
by Eq.\ref{eq:D-111} and Eq.\ref{eq:D-112}, we have

\begin{equation}
\sum_{j}\frac{2m\Delta v_{j}^{0}}{T_{j}^{0}}=0.\label{eq:D-114}
\end{equation}

By requiring the system to be a steady state, we can get $\sum_{j}\left(2m\Delta v_{j}^{0}/T_{j}^{0}\right)=0$.
However, it cannot tell us the $v_{j}^{0}-\mathrm{dependence}$ of
$\Delta v_{j}^{0}$, since we have summed over all $v_{j}^{0}$ and
thus lost such information. Such information can be figured out only
if we study individual $\Delta v_{j}^{0}$. This is one merit of the
mechanical model for calculating $\Delta v_{j}^{0}$ (Eq.\ref{eq:D-101}
and Fig. 9).

\subsection{Equation of state}

Now we are ready to write down the equation of state for a one-dimensional
gas with square well potential. Putting Eq.\ref{eq:B-23} and Eq.\ref{eq:D-114}
into Eq.\ref{eq:C-9} and keeping the physics to the lowest order
of particle-particle interaction, we have

\begin{equation}
F=\rho k_{B}T_{real}+\rho^{2}\left(k_{B}T_{real}\right)d_{1}-\rho^{2}\left(d_{2}-d_{1}\right)\epsilon.\label{eq:D-117}
\end{equation}
This is the equation of state for square well potential in our mechanical
approach, and it gives the same answer as statistical mechanics (Eq.\ref{eq:A-1}).

The physical meaning behind Eq.\ref{eq:D-117} is a little subtle.
Putting Eq.\ref{eq:B-23} and Eq.\ref{eq:D-114} into Eq.\ref{eq:2-8},
we have the equation of state of form

\begin{equation}
F=\rho k_{B}T+\rho^{2}\left(k_{B}T\right)d_{1}+\rho^{2}\left(d_{2}-d_{1}\right)\epsilon.\label{eq:D-118}
\end{equation}
As a limiting case, let us neglect the hard core and consider only
the effect of the particle-particle attraction. By Eq.\ref{eq:D-117}
and Eq.\ref{eq:D-118}, we have

\begin{equation}
\rho k_{B}T<F<\rho k_{B}T_{real}.\label{eq:D-119}
\end{equation}
Thus, to the naive question of ``When particle-particle attraction
is turned on, is the force increased or decreased?\textquotedblright{}
the answer is that the force is actually increased. But one should
follow by the immediate qualifier that, ``the temperature of the
system is also increased.\textquotedblright{} However, if we are comparing
the force exerted by an ideal gas and that by a van der Waals gas
\emph{at the same temperature}, but with the particle size effect
ignored, then the van der Waals gas necessarily exerts a smaller force.

\section{One-dimensional interacting gas with generic particle-particle interaction}

With the lessons learnt from one-dimensional interacting gas with
square well potential, we can generalize the mechanical model to generic
one-dimensional interacting gas. By generic particle-particle interaction,
we mean the potential is consisted of a hard core and an interaction
tail with finite range $r_{0}$. In the following, we study attraction
tail and then repulsion tail.

\subsection{Particle-particle attraction}

Obtain $\Delta t_{interaction}\left(0,v_{k}^{0}\right)$ by Eq.\ref{eq:A-5}
and then put it into Eq.\ref{eq:B-17}. Expand with $\frac{U(r)}{k_{B}T}$
to $\mathcal{O}\left(\frac{U(r)}{k_{B}T}\right)$. We have
\begin{equation}
\frac{1}{2}\Delta T_{j}^{0}=-\frac{Nd_{1}}{v_{j}^{0}}-\frac{N\sqrt{a\pi}}{\left(v_{j}^{0}\right)^{2}}e^{-\frac{\left(v_{j}^{0}\right)^{2}}{a}}\mathrm{erfi}(\frac{v_{j}^{0}}{\sqrt{a}})\left(\int_{d_{1}}^{r_{0}}dr\left[-\frac{U(r)}{k_{B}T}\right]\right).\label{eq:G-7}
\end{equation}
Eq.\ref{eq:G-7} is just Eq.\ref{eq:B-23}, with $\left(d_{2}-d_{1}\right)\frac{\epsilon}{k_{B}T}$
replaced by $\left(\int_{d_{1}}^{r_{0}}dr\left[-\frac{U(r)}{k_{B}T}\right]\right)$.
Therefore, the physics is all the same. For generic particle-particle
interaction being attractive, $U(r)$ is negative and particles are
sped up. Putting Eq.\ref{eq:D-114} and Eq.\ref{eq:G-7} into Eq.\ref{eq:C-9},
we obtain

\begin{equation}
F=\rho k_{B}T_{real}+\rho^{2}\left(k_{B}T_{real}\right)d_{1}-\rho^{2}\left(\int_{d_{1}}^{r_{0}}dr\left[-U(r)\right]\right).\label{eq:G-10}
\end{equation}
This is exactly the equation of state derived using standard statistical
mechanics, as we show in Eq.\ref{eq:1-3}.

The physics behind Eq.\ref{eq:G-10} is the same as what we present
in Section 3. Now we want to analyze the physics from another viewpoint
basing on the idea of average velocity. Instead of separating $\Delta T_{j}^{0}$
from $T_{j}=T_{j}^{0}+\Delta T_{j}^{0}$ (what we do from Eq.\ref{eq:2-7}
to Eq.\ref{eq:2-8}), we can regard $T_{j}$ as an entire entity and
then work with the idea of average velocity ($v_{j}\mid_{average}=2\left(L-Nd_{1}\right)/T_{j}$)
directly but not the free particle velocity. In such approach, ``physics
in the bulk'' is simply embedded in the idea of average velocity.
By our very construction, the only effect to be considered is what
happens when a particle hits the wall.

Rewrite the equation of state with average velocity and collision
velocity. Start from Eq.\ref{eq:2-7} and keep the physics to the
first order in the particle-particle interaction, and we have

\begin{equation}
F=\rho k_{B}T_{real}+\rho^{2}\left(k_{B}T_{real}\right)d_{1}+\frac{m}{L}\sum_{j}\left(v_{j}\mid_{average}\right)\left(\left(v_{j}^{0}+\Delta v_{j}^{0}\right)-v_{j}\mid_{average}\right).\label{eq:G-11}
\end{equation}
Compared with Eq.\ref{eq:G-10}, for particle-particle interaction
being attractive, the third term in Eq.\ref{eq:G-11} should be negative.
The reason hides in Table 1. From Table 1, we can see that, on the
average, a particle experiences more forward collisions from its neighbors
when it is in the bulk than it is on the boundary. Therefore, we have

\begin{equation}
\left(v_{j}^{0}+\Delta v_{j}^{0}\right)-v_{j}\mid_{average}<0.\label{eq:G-12}
\end{equation}

The feature that $\left(v_{j}^{0}+\Delta v_{j}^{0}\right)-v_{j}\mid_{average}<0$
can be seen in Fig. 9. So now we understand that it is caused by the
fact that the two kinds of velocities have different numbers of forward
collisions.

In a naive description trying to explain the form of the van der Waals
equation \cite{van2004continuity}, one is tempted to claim that a
particle hitting the wall is slowed down due to the attraction of
other gas particles from the back, and hence the force felt by the
wall is decreased. Put differently, this seemingly plausible theory
claims that the velocity hitting the wall (collision velocity) is
smaller than the velocity in the bulk (average velocity), which also
holds true in our detailed mechanical approach.

However, we must quickly point out that, as particle-particle attraction
is turned on, physics in the bulk is non-trivial and particles in
the box do fly faster. The reason that we don't need to consider average
velocity correction in the conventional description of the van der
Waals equation is not because forward collisions and backward collisions
cancel out exactly (they don't), but lies in the fact that they are
all lumped into the idea of $v_{j}\mid_{average}$. Now it is clear
that traditional picture of the van der Waals equation actually uses
$v_{j}\mid_{average}$ but not $v_{j}^{0}$ in the argument. For physics
in the bulk, it just takes the net result ($v_{j}\mid_{average}$)
and then bypasses all the issues related to ``what happens when a
particle flies in the middle of the box,\textquotedblright{} including
temperature modification. It is in this sense that traditional picture
makes sense.

\subsection{Particle-particle repulsion}

For particle-particle repulsion, we just reverse the physics of attraction
tail. However, one feature does stand out that is characteristic to
particle-particle repulsion alone. When the relative velocity of two
interacting particles is small, they don't have enough kinetic energy
(in center-of-mass frame) to overcome the potential barrier and reach
$r=d_{1}$. Nevertheless, the population of such particle pairs is
quite small since particle-particle interaction is weak by our construction,
and hence such effect is negligible in equation of state.

Another issue related to particle-particle repulsion is soft core.
An example of soft core is Lennard-Jones potential, which has a $r^{-12}$
repulsive term. Can we attack soft core in our mechanical approach?
The answer is a sadly no. Soft core aims to stop two colliding particles,
and this goal needs strong particle-particle interaction that cannot
be treated as a perturbation. Nonperturbative approach is needed for
studying soft core.

\section{Conclusion}

In this work, we have taken a mechanical approach to a one-dimensional
interacting gas with particle-particle interaction made up of a hard
core and an interaction tail. Perturbations around ideal gas are considered
and the physics is computed correct up to $\mathcal{O}\left(\frac{U}{k_{B}T}\right)$.
Equation of state given by our mechanical approach matches the result
given by statistical mechanics. Assumptions and conditions we put
in by hand include (1) the homogeneity assumption on the spatial distribution,
(2) velocity distribution being Maxwellian, (3) the system being in
steady state, (4) short range and weak interparticular interaction,
(5) low density, and (6) high temperature, all of which are common
assumed in traditional statistical mechanics.

\subsection{Summary}

Besides having been able to reproduce the equation of state, we do
obtain some physics insight that traditional statistical mechanics
doesn't readily tell us. In the following, we list six points for
the effects of particle-particle attraction. Almost all the physics
in the mechanical model is hidden in Fig. 9 and Table 1.

(1) Due to the fact that the number of forward collisions is larger
than the number of backward collisions, on the average, all particles
move faster in the presence of particle-particle attraction.

(2) Since particles move faster as particle-particle attraction is
turned on, the temperature of the system increases. Another point
of view without concerning mechanics is considering total energy of
the system. 

(3) For a slow particle, $\Delta v_{j}^{0}<0$. For a fast particle,
$\Delta v_{j}^{0}>0$. The increase and decrease of collision velocity
necessarily cancel out due to our very assumption that the system
is a steady state.

(4) Average velocity correction and collision velocity correction
have similar trends since the physics are similar. When the main particle
moves fast, the effect of forward collisions dominates and it leads
to the increasing trend. Suppression of the effect of particle-particle
interaction by exponential decay in Maxwell's velocity distribution
is responsible for the decreasing trend.

(5) Compared with collision velocity, average velocity has more forward
collisions, and hence average velocity is bigger than collision velocity.

(6) Compared with $\rho k_{B}T$, the force is increased. But compared
with $\rho k_{B}T_{real}$, the force is decreased.

\subsection{Meaning and implication}

Unlike most researches studying N-body systems in the framework of
statistical mechanics or kinetic theory, we take a mechanical approach
to a one-dimensional weakly interacting gas in thermal equilibrium.
With the help from a detailed analysis of how gas particles interact,
we have successfully gained some more interesting physics that the
traditional approach doesn't readily tell us. Our work is an example
demonstrating the possibility to study weakly interacting N-body systems
from a mechanical viewpoint working with particle trajectory.

Though the derivation here is focused on the equation of state, the
spirit of this mechanical model is investigating the physics of particle-particle
interaction and trying to have deeper understanding. With this mechanical
model, we obtain some understanding for issues other than the equation
of state.

How do we construct such mechanical model? There are three elements:
(1) a mechanical picture for unperturbed part (free particle model),
(2) plausible assumptions concerning things such as the velocity distribution
and homogeneity of the system that we put in by hand, and (3) assumed
form of (weak) particle-particle interaction. Adding the effect of
particle-particle interaction on free particle model, we capture some
interesting physics. The essence of such mechanical model lies in
the fact that the effect of particle-particle interaction is weak
($\mathcal{O}\left(\rho r_{0}\right)<\mathcal{O}\left(1\right)$ and
$\mathcal{O}\left(\frac{U}{k_{B}T}\right)<\mathcal{O}\left(1\right)$),
and so in some sense we can have the notion of particle trajectory
and focus on two-body interaction. For a system different from one-dimensional
interacting gas (some colloidal systems or granular systems, for example),
if we can figure out its free particle model and then put in particle-particle
interaction, maybe we can have some interesting physics. We think
it is possible to extend the idea to other systems that are usually
studied in a statistical mechanics or kinetic theory point of view.

Statistical mechanics works with partition function. Traditional kinetic
theory adopting Boltzmann equation considers a particle interacting
with its neighbor particles. Our mechanical approach works with a
more detailed analysis of the particle trajectory, the number of collisions
and the effect of each collision. The merit of this mechanical approach
is that it does provide us with more physics insight about how the
individual effect affects the final answer.

\pagebreak{}

\textbf{Appendix A: Probability of the last collision}\medskip{}

In this appendix, we calculate the probability of the last collision.
To find the probability of such situation, we put in the background
particles one by one. The probability of the background particle that
causes the last collision is given by

\begin{equation}
\left(\left\vert 1-\frac{v_{k}^{0}}{v_{j}^{0}}\right\vert \frac{\Delta d_{3}}{L}\right)\left(f\left(v_{k}^{0}\right)\Delta v_{k}^{0}\right).\label{eq:D-23-1}
\end{equation}

For the remaining $N-2$ background particles, they should be put
in the box in the manner that they cannot touch the main particle
after the main particle has its last collision with the first background
particle. The position of one of the remaining $N-2$ background particles
with free particle velocity $v_{l}^{0}$ at $t=\left(L-d_{3}\right)/v_{j}^{0}$
is denoted as $\widetilde{x_{l}}$. The requirement for $\widetilde{x_{l}}$
is

\begin{equation}
0<\widetilde{x_{l}}<L-d_{3},\label{eq:D-6-1}
\end{equation}
and the admissible velocity is

\begin{equation}
-\frac{L+\widetilde{x_{l}}}{d_{3}}v_{j}^{0}<v_{l}^{0}<\frac{L-\widetilde{x_{l}}}{d_{3}}v_{j}^{0}.\label{eq:D-7-1}
\end{equation}
The probability of a single background particle in the admissible
zone is thus given by
\begin{equation}
\int_{0}^{L-d_{3}}\frac{d\widetilde{x_{l}}}{L}\int_{-\frac{L+\widetilde{x_{l}}}{d_{3}}v_{j}^{0}}^{\frac{L-\widetilde{x_{l}}}{d_{3}}v_{j}^{0}}dv_{l}^{0}\frac{1}{\sqrt{a\pi}}e^{-\frac{\left(v_{l}^{0}\right)^{2}}{a}}.\label{eq:D-8-1}
\end{equation}
When thermodynamic limit is considered, we have
\begin{equation}
\mathrm{probability\ }\mathrm{of}\ (N-2)\ \mathrm{background}\ \mathrm{particles}=\mathrm{exp}\left\{ -\frac{1}{2}\left[1+\mathrm{erf}\left(\frac{v_{j}^{0}}{\sqrt{a}}\right)+\frac{\sqrt{a}}{\sqrt{\pi}v_{j}^{0}}e^{-\frac{\left(v_{j}^{0}\right)^{2}}{a}}\right]\rho d_{3}\right\} .\label{eq:D-16-2}
\end{equation}
The complete probability of the last collision is the product of Eq.\ref{eq:D-23-1}
and Eq.\ref{eq:D-16-2} and the permutation factor $N$, and is given
by
\begin{equation}
N\left(\left\vert 1-\frac{v_{k}^{0}}{v_{j}^{0}}\right\vert \frac{\Delta d_{3}}{L}\right)\left(f\left(v_{k}^{0}\right)\Delta v_{k}^{0}\right)\mathrm{exp}\left\{ -\frac{1}{2}\left[1+\mathrm{erf}\left(\frac{v_{j}^{0}}{\sqrt{a}}\right)+\frac{\sqrt{a}}{\sqrt{\pi}v_{j}^{0}}e^{-\frac{\left(v_{j}^{0}\right)^{2}}{a}}\right]\rho d_{3}\right\} .\label{eq:D-24-1}
\end{equation}

\pagebreak{}

\textbf{Appendix B: Physics on the boundary for forward collision}\medskip{}

In this appendix, we study the range of $d_{3}$ and the collision
velocity for forward collision. By Eq.\ref{eq:A-2} and Eq.\ref{eq:D-25},
the condition for $d_{3}$ is

\begin{equation}
d_{3}<d_{2}+\left(d_{2}-d_{1}\right)\frac{\frac{v_{j}^{0}+v_{k}^{0}}{2}}{\sqrt{\left(\frac{v_{j}^{0}-v_{k}^{0}}{2}\right)^{2}+\frac{\epsilon}{m}}}.\label{eq:D-28-1}
\end{equation}
If the main particle hits the wall when it is in the potential well,
by Eq.\ref{eq:D-25}, the collision velocity is

\begin{equation}
v_{j}^{0}+\Delta v_{j}^{0}=\frac{v_{j}^{0}+v_{k}^{0}}{2}+\sqrt{\left(\frac{v_{j}^{0}-v_{k}^{0}}{2}\right)^{2}+\frac{\epsilon}{m}}.\label{eq:D-27-1}
\end{equation}
By Eq.\ref{eq:D-26}, Eq.\ref{eq:D-28-1} and Eq.\ref{eq:D-27-1},
the nonvanishing $\Delta v_{j}^{0}$ in forward collision part is
summarized in Fig. 10, with

\begin{figure}
\begin{centering}
\includegraphics[scale=0.45]{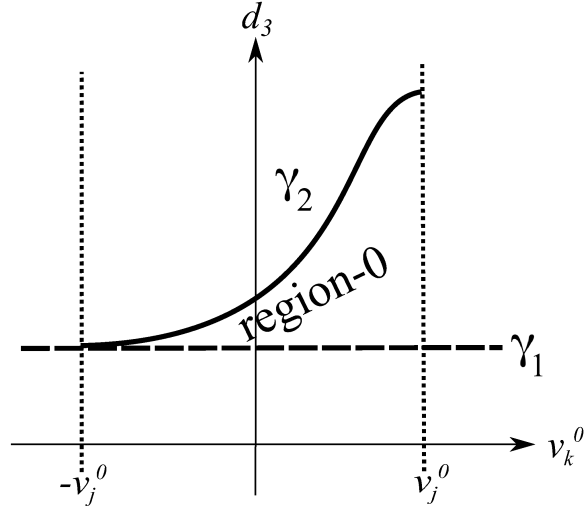}
\par\end{centering}
\caption{Nonvanishing $\Delta v_{j}^{0}$ in forward collision part. Nonvanishing
$\Delta v_{j}^{0}$ appears in region-0. Detailed information is given
by Eq.\ref{eq:D-29-1} and Eq.\ref{eq:D-30-1}.}
\end{figure}

\begin{eqnarray}
\gamma_{1} & : & d_{3}=d_{2},\nonumber \\
\gamma_{2} & : & d_{3}=d_{2}+\left(d_{2}-d_{1}\right)\frac{\frac{v_{j}^{0}+v_{k}^{0}}{2}}{\sqrt{\left(\frac{v_{j}^{0}-v_{k}^{0}}{2}\right)^{2}+\frac{\epsilon}{m}}},\label{eq:D-29-1}
\end{eqnarray}

\begin{equation}
\Delta v_{j}^{0}\mid_{region-0}=\frac{-v_{j}^{0}+v_{k}^{0}}{2}+\sqrt{\left(\frac{v_{j}^{0}-v_{k}^{0}}{2}\right)^{2}+\frac{\epsilon}{m}}.\label{eq:D-30-1}
\end{equation}

\pagebreak{}

\textbf{Appendix C: Physics on the boundary for backward collision}\medskip{}

In this appendix, we study the range of $d_{3}$ and the collision
velocity for backward collision. By Eq.\ref{eq:A-2} and Eq.\ref{eq:D-25},
the condition for $d_{3}$ is
\begin{equation}
\left(d_{2}+d_{3}\right)<d_{2}+\left(d_{2}-d_{1}\right)\frac{\frac{v_{j}^{0}+v_{k}^{0}}{2}}{\sqrt{\left(\frac{v_{j}^{0}-v_{k}^{0}}{2}\right)^{2}+\frac{\epsilon}{m}}}.\label{eq:D-32-1}
\end{equation}
Eq.\ref{eq:D-31} and Eq.\ref{eq:D-32-1} are necessary conditions
to have nonvanishing $\Delta v_{j}^{0}$. Under the two necessary
conditions, the situations can be classified as three classes (Class
1, Class 2 and Class 3) by different forms of collision velocities.

\begin{figure}
\begin{centering}
\includegraphics[scale=0.6]{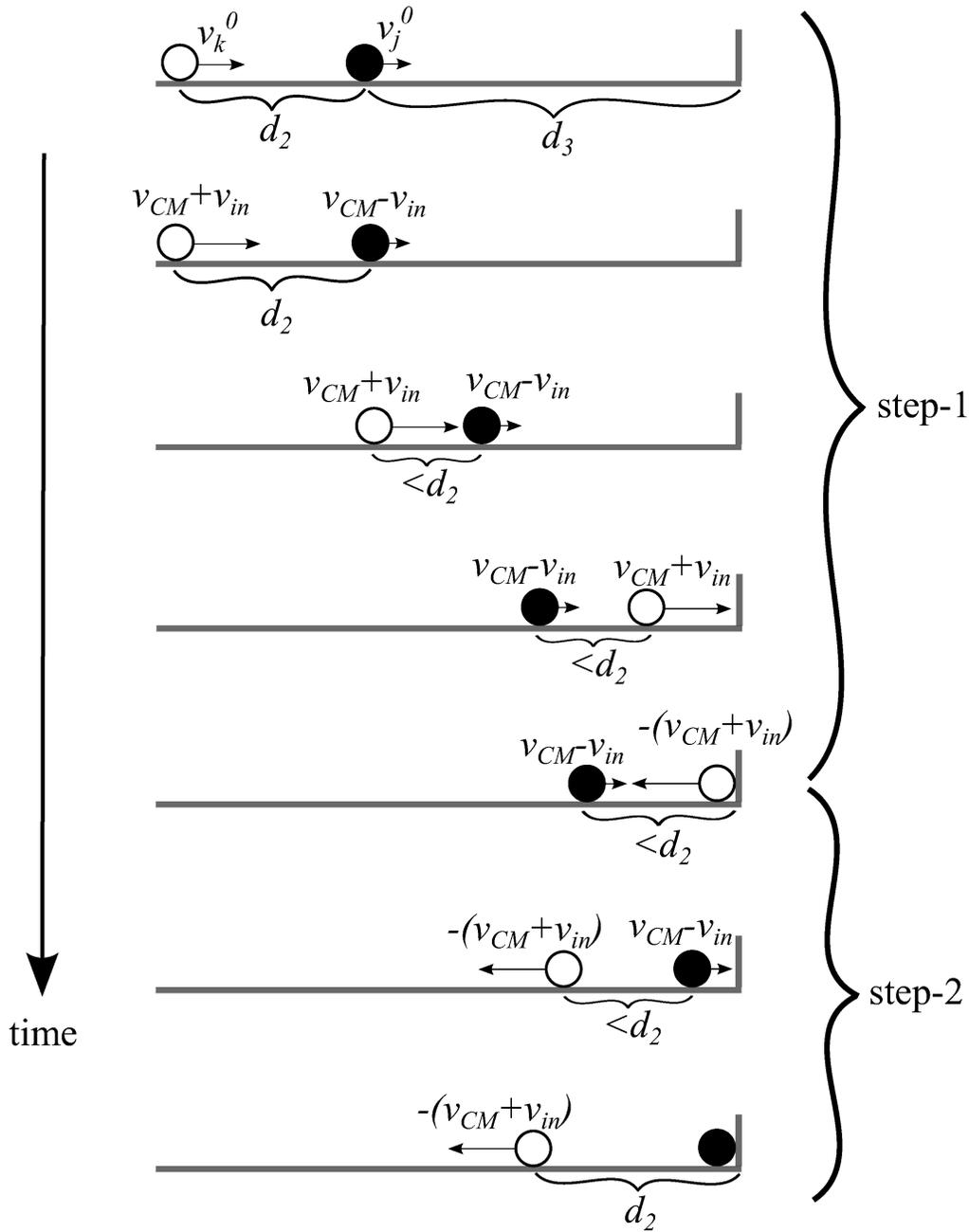}
\par\end{centering}
\caption{Critical situation that separates the three classes (Class 1, Class
2 and Class 3). The figure shows the motion of the main particle and
the first background particle in the lab frame. Step-1 starts with
the beginning of backward collision and ends with the event that the
first background particle hits the wall. Step-2 starts with the end
of step-1 and ends with the moment that the two particles are going
to decouple. The value of $d_{3}$ in this critical situation is denoted
as $d_{3}\mid_{critical}$.}
\end{figure}

The critical situation separating the three classes is shown in Fig.
11, with $d_{3}\mid_{critical}$ given by

\begin{equation}
d_{3}\mid_{critical}=\left(d_{2}-d_{1}\right)\left(\frac{\frac{v_{j}^{0}+v_{k}^{0}}{2}}{\sqrt{\left(\frac{v_{j}^{0}-v_{k}^{0}}{2}\right)^{2}+\frac{\epsilon}{m}}}-1\right).\label{eq:D-37-1}
\end{equation}

Class 1:\\
In Class 1, $d_{3}<d_{3}\mid_{critical}$. In this class, the main
particle hits the wall when it is in the potential well. The collision
velocity is

\begin{equation}
v_{j}^{0}+\Delta v_{j}^{0}=\frac{v_{j}^{0}+v_{k}^{0}}{2}-\sqrt{\left(\frac{v_{j}^{0}-v_{k}^{0}}{2}\right)^{2}+\frac{\epsilon}{m}}.\label{eq:D-38-1}
\end{equation}

Class 2:\\
In Class 2, $d_{3}>d_{3}\mid_{critical}$. In this class, the main
particle hits the wall after it decouples with the first background
particle and becomes a free particle again. However, the velocity
of the main particle being a free particle is no longer $v_{j}^{0}$,
since the wall comes into play and alters particle-particle interaction
via particle-wall collision of the first background particle at the
end of step-1. With the help of Eq.\ref{eq:D-25}, the new free particle
velocity of the main particle, namely the collision velocity, is given
by

\begin{equation}
v_{j}^{0}+\Delta v_{j}^{0}=\sqrt{\left(\frac{v_{j}^{0}+v_{k}^{0}}{2}\right)^{2}-\frac{\epsilon}{m}}-\sqrt{\left(\frac{v_{j}^{0}-v_{k}^{0}}{2}\right)^{2}+\frac{\epsilon}{m}}.\label{eq:D-42-1}
\end{equation}

Class 3:\\
In Fig. 11, we actually already assume that the main particle always
has a positive velocity. But this is not always guaranteed. If the
main particle has a negative velocity, it cannot hit the wall, and
the collision velocity in such case is zero. There are two situations
when no collision happens between the main particle and the wall.

The first situation is that the velocity of the main particle is negative
at the moment right after the collision starts. That is, $v_{CM}-v_{in}<0$.
In this case, $v_{k}^{0}$ is given by

\begin{equation}
v_{k}^{0}<\frac{\frac{\epsilon}{m}}{v_{j}^{0}}.\label{eq:D-45-1}
\end{equation}
The second situation is that the velocity of the main particle is
negative after they decouple. That is, the new free particle velocity
in Eq.\ref{eq:D-42-1} is negative. In this case, $v_{k}^{0}$ is
given by

\begin{equation}
\frac{\frac{\epsilon}{m}}{v_{j}^{0}}<v_{k}^{0}<\frac{2\frac{\epsilon}{m}}{v_{j}^{0}}.\label{eq:D-47-1}
\end{equation}

In Class 1, the main particle hits the wall during its interaction
with the first background particle. In Class 2, the main particle
hits the wall when it becomes a free particle after it decouples with
the first background particle. In Class 3, the main particle simply
cannot hit the wall.

\begin{figure}
\begin{centering}
\includegraphics[scale=0.35]{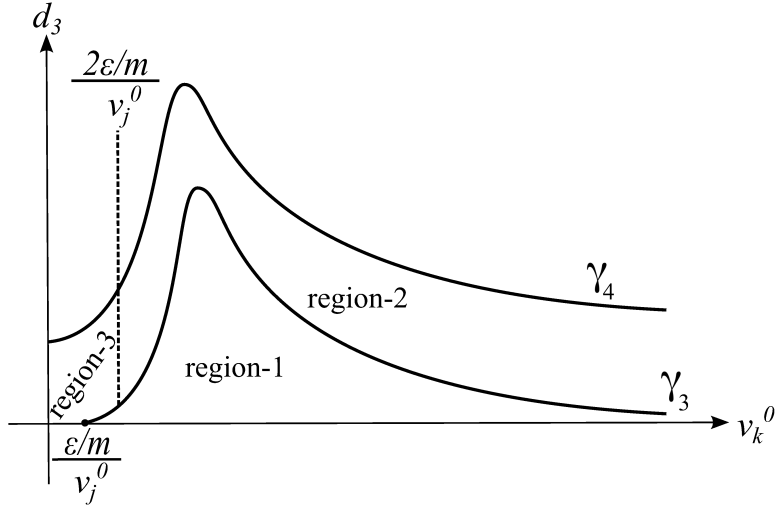}
\par\end{centering}
\caption{Nonvanishing $\Delta v_{j}^{0}$ in backward collision part. Region-1,2,3
corresponds to Class 1,2,3, respectively. In addition, region for
$v_{k}^{0}<v_{j}^{0}$ is forbidden. Detailed information is given
by Eq.\ref{eq:D-48-1} and Eq.\ref{eq:D-49-1}.}

\end{figure}

The nonvanishing $\Delta v_{j}^{0}$ in backward collision part is
summarized in Fig. 12, with
\begin{eqnarray}
\gamma_{3} & : & d_{3}=\left(d_{2}-d_{1}\right)\left(\frac{\frac{v_{j}^{0}+v_{k}^{0}}{2}}{\sqrt{\left(\frac{v_{j}^{0}-v_{k}^{0}}{2}\right)^{2}+\frac{\epsilon}{m}}}-1\right),\nonumber \\
\gamma_{4} & : & d_{3}=\left(d_{2}-d_{1}\right)\frac{\frac{v_{j}^{0}+v_{k}^{0}}{2}}{\sqrt{\left(\frac{v_{j}^{0}-v_{k}^{0}}{2}\right)^{2}+\frac{\epsilon}{m}}},\label{eq:D-48-1}
\end{eqnarray}

\begin{eqnarray}
\mathrm{Class}\ 1 & : & \Delta v_{j}^{0}\mid_{region-1}=\frac{-v_{j}^{0}+v_{k}^{0}}{2}-\sqrt{\left(\frac{v_{j}^{0}-v_{k}^{0}}{2}\right)^{2}+\frac{\epsilon}{m}},\nonumber \\
\mathrm{Class}\ 2 & : & \Delta v_{j}^{0}\mid_{region-2}=-v_{j}^{0}+\sqrt{\left(\frac{v_{j}^{0}+v_{k}^{0}}{2}\right)^{2}-\frac{\epsilon}{m}}-\sqrt{\left(\frac{v_{j}^{0}-v_{k}^{0}}{2}\right)^{2}+\frac{\epsilon}{m}},\nonumber \\
\mathrm{Class}\ 3 & : & \Delta v_{j}^{0}\mid_{region-3}=-v_{j}^{0}.\label{eq:D-49-1}
\end{eqnarray}
Of course, we also have the condition of Eq.\ref{eq:D-31}, and hence
the region that $v_{k}^{0}<v_{j}^{0}$ is forbidden. Therefore, Class
3 in backward collision part occurs only for small $u_{j}^{0}$ satisfying

\begin{equation}
u_{j}^{0}<\sqrt{\frac{\epsilon}{k_{B}T}}.\label{eq:D-51-1}
\end{equation}

\medskip{}

\medskip{}

\medskip{}

\medskip{}

\medskip{}

\textbf{Appendix D: Total effect of the correction of collision velocity}\medskip{}

In this appendix, we finally calculate the total effect of the correction
of collision velocity. The task we are going to do is putting Eq.\ref{eq:D-24}
(probability), Eq.\ref{eq:D-30-1} ($\Delta v_{j}^{0}$ for forward
collision part) and Eq.\ref{eq:D-49-1} ($\Delta v_{j}^{0}$ for backward
collision part) together and then integrate over the four regions
shown in Fig. 10 and Fig. 12. That is, we are going to deal with the
integral
\begin{eqnarray}
\Delta v_{j}^{0} & = & \rho\int d(d_{3})\int dv_{k}^{0}f\left(v_{k}^{0}\right)\left\vert 1-\frac{v_{k}^{0}}{v_{j}^{0}}\right\vert \mathrm{exp}\left\{ -\frac{1}{2}\left[1+\mathrm{erf}\left(\frac{v_{j}^{0}}{\sqrt{a}}\right)+\frac{\sqrt{a}}{\sqrt{\pi}v_{j}^{0}}e^{-\frac{\left(v_{j}^{0}\right)^{2}}{a}}\right]\rho d_{3}\right\} \nonumber \\
 &  & \times\left(\Delta v_{j}^{0}\mid_{region-0}+\Delta v_{j}^{0}\mid_{region-1}+\Delta v_{j}^{0}\mid_{region-2}+\Delta v_{j}^{0}\mid_{region-3}\right).\label{eq:D-52-1}
\end{eqnarray}

Let's stop and think whether Eq.\ref{eq:D-52-1} can be simplified
further. The value of $d_{3}$ in region-0,1,2,3 is quite small. Therefore,
it is all right to replace the exponential decay part in the probability
distribution by unity, due to the feature that the probability decays
slowly. The argument is in Appendix E. Furthermore, since $\Delta v_{j}^{0}\mid_{region-3}$
exists only for small velocity (Eq.\ref{eq:D-51-1}), the population
is small and hence the contribution is negligible for equation of
state. The argument is in Appendix F. After the two simplifications,
Eq.\ref{eq:D-52-1} is now simplified as
\begin{equation}
\Delta v_{j}^{0}=\rho\int d(d_{3})\int dv_{k}^{0}f\left(v_{k}^{0}\right)\left\vert 1-\frac{v_{k}^{0}}{v_{j}^{0}}\right\vert \left(\Delta v_{j}^{0}\mid_{region-0}+\Delta v_{j}^{0}\mid_{region-1}+\Delta v_{j}^{0}\mid_{region-2}\right).\label{eq:D-91-1}
\end{equation}
Calculate $\Delta v_{j}^{0}$ via Eq.\ref{eq:D-91-1} and expand the
result with $\frac{\epsilon}{k_{B}T}$ to $\mathcal{O}\left(\frac{\epsilon}{k_{B}T}\right)$.
We get

\begin{equation}
\Delta u_{j}^{0}=\frac{\Delta v_{j}^{0}}{\sqrt{a}}=\frac{\rho\left(d_{2}-d_{1}\right)}{2\sqrt{\pi}u_{j}^{0}}\left(\int_{-u_{j}^{0}}^{\infty}du_{k}^{0}e^{-\left(u_{k}^{0}\right)^{2}}\frac{u_{j}^{0}+u_{k}^{0}}{u_{j}^{0}-u_{k}^{0}}+\int_{u_{j}^{0}}^{\infty}du_{k}^{0}e^{-\left(u_{k}^{0}\right)^{2}}\frac{u_{j}^{0}-u_{k}^{0}}{u_{j}^{0}+u_{k}^{0}}\right)\frac{\epsilon}{k_{B}T}.\label{eq:D-99-1}
\end{equation}

\medskip{}

\medskip{}
\medskip{}
\medskip{}

\medskip{}

\medskip{}

\medskip{}

\medskip{}
\medskip{}

\medskip{}

\medskip{}

\textbf{Appendix E: Simplification of the exponential decay probability
distribution}\medskip{}

In this appendix, we argue that the exponential decay probability
distribution in Eq.\ref{eq:D-52-1} can be replaced by unity. The
basic idea is trying to show that $\mathrm{exp}\left\{ -\frac{1}{2}\left[1+\mathrm{erf}\left(\frac{v_{j}^{0}}{\sqrt{a}}\right)+\frac{\sqrt{a}}{\sqrt{\pi}v_{j}^{0}}e^{-\frac{\left(v_{j}^{0}\right)^{2}}{a}}\right]\rho d_{3}\right\} $
decays with $d_{3}$ in a slow manner. To deal with the idea of ``decays
in a slow manner,'' we define the ``decay depth'' as

\begin{equation}
\mathrm{decay\ depth}\equiv\left(\frac{1}{2}\left[1+\mathrm{erf}\left(\frac{v_{j}^{0}}{\sqrt{a}}\right)+\frac{\sqrt{a}}{\sqrt{\pi}v_{j}^{0}}e^{-\frac{\left(v_{j}^{0}\right)^{2}}{a}}\right]\rho\right)^{-1}.\label{eq:D-53-1}
\end{equation}
On the other hand, from Fig. 10 and Fig. 12, we can see that

\begin{equation}
\mathrm{maximum\ of\ }\mathrm{relevant\ }d_{3}=d_{2}+\left(d_{2}-d_{1}\right)\frac{\frac{u_{j}^{0}+u_{k}^{0}}{2}}{\sqrt{\left(\frac{u_{j}^{0}-u_{k}^{0}}{2}\right)^{2}+\frac{\epsilon}{2k_{B}T}}}.\label{eq:D-57-1}
\end{equation}
So we have

\begin{equation}
\mathcal{O}\left(\frac{\mathrm{maximum\ of\ relevant}\ d_{3}}{\mathrm{decay\ depth}}\right)=\mathcal{O}\left(\rho\left[1+\frac{1}{u_{j}^{0}}e^{-\left(u_{j}^{0}\right)^{2}}\right]\left(d_{2}+\left(d_{2}-d_{1}\right)\frac{\frac{u_{j}^{0}+u_{k}^{0}}{2}}{\sqrt{\left(\frac{u_{j}^{0}-u_{k}^{0}}{2}\right)^{2}+\frac{\epsilon}{2k_{B}T}}}\right)\right).\label{eq:D-58-1}
\end{equation}

We want $\mathcal{O}\left(\frac{\mathrm{maximum\ of\ relevant}\ d_{3}}{\mathrm{decay\ depth}}\right)<\mathcal{O}\left(1\right)$.
If $\mathcal{O}\left(\frac{\mathrm{maximum\ of\ relevant}\ d_{3}}{\mathrm{decay\ depth}}\right)$
is not smaller than unity, we have a trouble. With some analysis for
Eq.\ref{eq:D-58-1}, we can find that there are some $\left(u_{j}^{0},u_{k}^{0}\right)$
that don't respect $\mathcal{O}\left(\frac{\mathrm{maximum\ of\ relevant}\ d_{3}}{\mathrm{decay\ depth}}\right)<\mathcal{O}\left(1\right)$.
However, the violations appear for small $u_{j}^{0}$ or the case
that $u_{k}^{0}$ is very close to $u_{j}^{0}$. The result is, the
contributions given by these trouble makers are negligible in the
order being considered, due to the fact that their populations are
small.

\medskip{}

\medskip{}
\medskip{}

\medskip{}

\medskip{}

\textbf{Appendix F: Contribution of $\Delta v_{j}^{0}\mid_{region-3}$}\medskip{}

In this appendix, we argue that the contribution of $\Delta v_{j}^{0}\mid_{region-3}$
is negligible when we sum over all velocities to get the correction
in momentum transferred. For $u_{j}^{0}<\sqrt{\frac{\epsilon}{k_{B}T}}$,
we have $\Delta v_{j}^{0}\mid_{region-3}=-v_{j}^{0}$. So we have
\begin{eqnarray}
 &  & \left[\sum_{j}\frac{2m\Delta v_{j}^{0}}{T_{j}^{0}}\right]_{u_{j}^{0}<\sqrt{\frac{\epsilon}{k_{B}T}},\mathrm{Class}\ 3}\nonumber \\
 & = & -\frac{4}{\pi}\rho^{2}k_{B}T\left[\int_{0}^{\sqrt{\frac{\epsilon}{k_{B}T}}}du_{j}^{0}e^{-\left(u_{j}^{0}\right)^{2}}\left(u_{j}^{0}\right)^{2}\int_{u_{j}^{0}}^{\frac{\epsilon}{k_{B}T}\frac{1}{u_{j}^{0}}}du_{k}^{0}e^{-\left(u_{k}^{0}\right)^{2}}\left(1-\frac{u_{k}^{0}}{u_{j}^{0}}\right)\int d(d_{3})\right].\label{eq:D-85-1}
\end{eqnarray}
By Eq.\ref{eq:D-48-1} and Fig. 12, the integral of $d_{3}$ is bounded
by

\begin{equation}
\int d(d_{3})<\left(d_{2}-d_{1}\right)\frac{u_{j}^{0}+u_{k}^{0}}{\sqrt{\left(u_{j}^{0}-u_{k}^{0}\right)^{2}+2\frac{\epsilon}{k_{B}T}}}.\label{eq:D-86-1}
\end{equation}
And so we have
\begin{eqnarray}
 &  & \frac{\pi}{4}\left\vert \left[\sum_{j}\frac{2m\Delta v_{j}^{0}}{T_{j}^{0}}\right]_{u_{j}^{0}<\sqrt{\frac{\epsilon}{k_{B}T}},\mathrm{Class}\ 3}\right\vert \nonumber \\
 & < & \rho^{2}k_{B}T\left(d_{2}-d_{1}\right)\left[\int_{0}^{\sqrt{\frac{\epsilon}{k_{B}T}}}du_{j}^{0}e^{-\left(u_{j}^{0}\right)^{2}}\left(u_{j}^{0}\right)^{2}\int_{u_{j}^{0}}^{\frac{\epsilon}{k_{B}T}\frac{1}{u_{j}^{0}}}du_{k}^{0}e^{-\left(u_{k}^{0}\right)^{2}}\right]\nonumber \\
 &  & +\rho^{2}k_{B}T\left(d_{2}-d_{1}\right)\left[\int_{0}^{\sqrt{\frac{\epsilon}{k_{B}T}}}du_{j}^{0}e^{-\left(u_{j}^{0}\right)^{2}}u_{j}^{0}\int_{u_{j}^{0}}^{\frac{\epsilon}{k_{B}T}\frac{1}{u_{j}^{0}}}du_{k}^{0}e^{-\left(u_{k}^{0}\right)^{2}}u_{k}^{0}\right].\label{eq:D-87-1}
\end{eqnarray}
The first term and the second term in the right hand side of Eq.\ref{eq:D-87-1}
are both bounded by $\mathcal{O}\left(\left[\rho^{2}\left(d_{2}-d_{1}\right)\epsilon\right]\sqrt{\frac{\epsilon}{k_{B}T}}\right)$
and thus are negligible.

\bibliographystyle{unsrt}
\bibliography{paper_reference}

\end{document}